\documentclass[%
 reprint,
superscriptaddress,
 amsmath,amssymb,
prx,
showkeys
]{revtex4-2}

\usepackage{graphicx}
\usepackage{dcolumn}
\usepackage{bm}

\begin{document}

\preprint{APS/123-QED}

\title{Energy scaling of water window high-order harmonic generation for \\single-shot soft X-ray spectroscopy and live-cell imaging}

\author{Yuxi Fu}
\email{These authors contributed equally to this work.}
 \affiliation{%
 Attosecond Science Research Team, RIKEN Center for Advanced Photonics, RIKEN, 2-1 Hirosawa, Wako, Saitama 351-0198, Japan.
}%

\author{Kotaro Nishimura}%
\email{These authors contributed equally to this work.}
 \affiliation{%
 Attosecond Science Research Team, RIKEN Center for Advanced Photonics, RIKEN, 2-1 Hirosawa, Wako, Saitama 351-0198, Japan.
}%
\affiliation{%
Department of Physics, Tokyo University of Science, 2641 Yamazaki, Noda, Chiba 278-8510, Japan.
}%

 \author{Renzhi Shao}
 \affiliation{%
School of Physics and Wuhan National Laboratory for Optoelectronics, Huazhong University of Science and Technology, Wuhan 430074, China.}%

 \author{Akira Suda}
\affiliation{%
Department of Physics, Tokyo University of Science, 2641 Yamazaki, Noda, Chiba 278-8510, Japan.
}%

\author{Katsumi Midorikawa}
\affiliation{%
 Attosecond Science Research Team, RIKEN Center for Advanced Photonics, RIKEN, 2-1 Hirosawa, Wako, Saitama 351-0198, Japan.
}%
 
\author{Pengfei Lan}
\affiliation{%
School of Physics and Wuhan National Laboratory for Optoelectronics, Huazhong University of Science and Technology, Wuhan 430074, China.}%

\author{Eiji J. Takahashi}
 \email{Corresponding author: ejtak@riken.jp}
\affiliation{%
 Attosecond Science Research Team, RIKEN Center for Advanced Photonics, RIKEN, 2-1 Hirosawa, Wako, Saitama 351-0198, Japan.
}%

\date{\today}

\begin{abstract}
 Full coherent soft X-ray attosecond pulses are now available through high-order harmonic generation (HHG); however, its insufficient output energy hinders various applications, such as attosecond-scale soft X-ray nonlinear experiments, the seeding of soft X-ray free-electron lasers, attosecond-pump-attosecond-probe spectroscopies, and single-shot imaging. 
In this paper, towards the implementation of these exciting studies, we demonstrate a soft X-ray harmonic beam that is more than two orders of magnitudes stronger up to the water window region compared to previous works. 
This was achieved by combining a newly developed TW class mid-infrared femtosecond laser and a loosely focusing geometry for HHG in the mid-infrared region for the first time.
Thanks to a loosely focusing geometry with a neutral medium target, we achieve a high conversion efficiency, a low beam divergence, and a significantly reduced medium gas pressure.
As the first application of our nano-joule level water window soft X-ray harmonic source, we demonstrate near edge X-ray absorption fine structure (NEXAFS) experiments with clear fine absorption spectra near the K- and L-edges observed in various samples.
The systematic study of a robust energy scaling method on HHG opens the door for demonstrating single-shot absorption spectrum and live cell imaging with a femtosecond time resolution.

%
\end{abstract}

\maketitle


\section{Introduction}

After the first observation of high-order harmonic generation (HHG) \cite{FirstHHG}, a compact XUV/soft X-ray laser source with attosecond ($10^{-18}$s) pulse duration and full coherence became feasible \cite{FirstAtto}, which has enabled the investigation of various ultrafast dynamics. 
One interesting and important research topic is nonlinear phenomena via attosecond time resolution, for example, the two-photon absorption process, which has only been investigated in the extreme ultraviolet (XUV) region using isolated attosecond pulses (IAPs) \cite{TakNC} and attosecond pulse trains (APTs) \cite{NabekawaPRL}.
 In fact, X-ray pulses are more useful for two-photon absorption owing to the much deeper penetration depth and more element-specific features \cite{XFELDoumy}.
However, the considerable reduced two-photon absorption cross section with respect to the photon energy requires a higher pulse energy from the X-ray pulses generated from HHG, which motivated us to develop a strong soft X-ray attosecond high-order harmonic (HH) source.

  Alternatively, X-ray free-electron laser (XFEL) facilities can provide high pulse energy soft X-rays, but its pulse duration has been normally limited to tens femtosecond \cite{XFELDoumy}.
    Very recently, enen though 100-as order isolated pulse has been reported by modified XFEL system \cite{XFELDuris}, the pulse energy and spectrum stability need to be further improved. 
  In general, as a consequence of the self-amplified spontaneous emission (SASE) scheme, the X-ray beam from XFELs has a poor temporal coherence and high pulse-to-pulse fluctuation.
  To improve the temporal coherence of FELs, a seeding scheme combined with HHG has been proposed.
  In 2011, \textit{Togashi}, \textit{Takahashi}, and co-workers demonstrated a seeded XUV-FEL using a XUV HHs seeder pulse\cite{TogashiOE}.
  They achieved 650 times the intensity compared to the unseeded condition and dramatically reduced the shot-to-shot fluctuation.
  In fact, this seeding approach can be extended to the soft X-ray region, e.g., the water window region (284-543 eV), which requires a threshold seeding pulse energy on the nJ level.
  Thus, a nanojoule class coherent soft X-ray source has been needed as a XFEL seeded pulse.

   Another important application is attosecond-pump-attosecond-probe experiments in the soft X-ray region, for example, time-resolved X-ray absorption fine structure (XAFS) research of the two-photon nonlinear process.
   Moreover, further promoting various applications, such as single-shot high-resolution imaging \cite{RavasioPRL}, optical surface contamination, mask defect inspection \cite{NagataMask}, and nanoscale transient gratings excitation \cite{BencivengaSciAdv}, requires a high-energy coherent soft X-ray source.
   In recent years, rapidly developed high-quality X-ray sources based on HHG have opened the door for applications on an attosecond time scale \cite{BuadesOptica}; however, the insufficient energy has restricted a variety of applications.

Instead of a well-developed and commercially available near-infrared (NIR) Ti:sapphire laser system, a mid-infrared (MIR) \cite{DiMauroNatPhys} femtosecond laser is highly desirable for efficiently generating soft X-ray HHs. This is because the maximum photon energy of HHs is not only approximately proportional to the laser intensity but is more prominently proportional to the driving laser wavelength squared \cite{Corkum3step}.
Thus, a MIR laser can effectively extend the HH photon energy to the soft X-ray region without strongly ionizing the generation medium, which can dramatically destroy the phase matching between the HH and driving laser \cite{TakWW,ChenPRL}.
However, the conversion efficiency of HHG decreases with a longer driving laser wavelength with $\lambda$\textsuperscript{-(5-6)}. 
To obtain a high-energy harmonic beam, a high-energy femtosecond MIR laser is needed to make up for the lower conversion efficiency of the longer driver wavelength.
In recent years, many studies have been devoted to generate soft X-ray HHs, especially in the water window region, driven by MIR femtosecond laser sources \cite{RothardtOL2014,IshiiNC,SteinJPB,TeichmannNC,LiNC,CardinJPB}.
However, the limited driving laser energy, which is typically in the mJ or sub-mJ range, has restricted the pulse energy of soft X-ray HHs to the pJ and even fJ level. Very recently, Johnson and co-workers increased the pulse energy to 71 pJ in the water window region by employing a sub-mJ 2-cycle MIR laser under an $over~driven$ scheme that depends on non-adiabatic phenomenon\cite{JohnsonSciAdv}.

In this paper, we demonstrate a robust approach to generate over nJ/pulse energy HHs in the water window soft X-ray, using our recently developed TW-class MIR femtosecond laser system with a maximum pulse energy of 100 mJ in the wavelength tuning range of 1.2-2.4 $\mu$m \cite{FuSR,FuJSTQE} using a dual-chirped optical parametric amplification (DC-OPA) scheme \cite{FuOL}.
By combining a meter-long loose focusing geometry, a neutral gas medium, and a low gas pressure, we achieved optimum phase matching and high conversion efficiency \cite{TakAPL} in the soft X-ray region.
From 150 to 300 eV, we obtained a 10 nJ pulse energy with a peak power that can reach the MW range, assuming the pulse has a temporal envelope on the order of 10 fs.
 In the water window region, we obtained over 3.5 nJ of pulse energy, which is approximately two orders higher than those for other reported HHs thus far.
 In our simulation, our method showed several advantages compared with the $over~driven$ scheme. 
 It is straightforward for energy scaling, reduced the requirement for the driving laser quality both spatially and temporally, involved a much lower gas pressure, and involved an easily predictable atto-chirp without complicated simulations.
 In the first application, we applied the HH source to X-ray absorption spectroscopy for various samples, and clear fine structures near the K-edges of carbon and L-edges of chlorine were observed.
 The experimental results indicate the  feasibility of quick measurements including single-shot of absorption spectra by using our developed water window HH source.

\section{Results}
%

\begin{figure}[htbp]
\centering
\includegraphics[width=\linewidth]{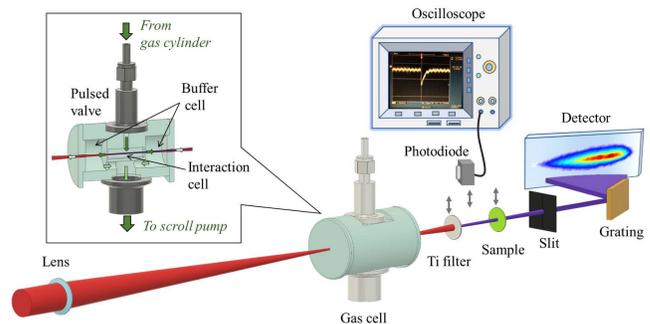}
\caption{\label{fig:setup} Schematic of the experimental setup.}
\end{figure}

The schematic of the experimental setup can be found in Fig. \ref{fig:setup}. 
The driving laser for HHG is provided by a DC-OPA system, which is pumped by a 1-J-class Ti:sapphire-chirped pulse amplification (CPA) system. The DC-OPA system can provide wavelength tunable laser pulses from 1.2 to 4 $\mu$m with pulses compressed to close to TL durations of 1.2-1.6 $\mu$m and 3-4 $\mu$m with TW class peak power.
In this experiment, the central wavelength of the DC-OPA was tuned to 1.55 $\mu$m, which is an air transmission window, with a duration of 45 fs (FWHM) and a pulse energy of up to 80 mJ \cite{FuSR}.
The high-energy pulses were loosely focused by an $f$ = 2 m lens (CaF\textsubscript{2}) with an anti-reflection coating.
The pulse energy, beam size and focus spot were adjusted by truncating the beam size using a variable aperture.
The pulse energies employed for different gas media under 2-m focusing were 10--15 mJ for Ar, 20--30 mJ for Ne, and 30--50 mJ for He.

The gas cell was designed as a double structure, of which the inner cell  can be filled with gas, whereas the outer cell  is employed for differential pumping to maintain the vacuum inside the chamber. The outer cell is pumped by a fast scroll pump with a pump capacity of 5000 L/min.
The inner cell is provided by a copper tube of various diameters.
The pinholes of the copper tube are drilled by the MIR laser pulses.
However, for He gas medium, the gas cell, which is a 3-cm-long tube with both ends sealed by thin cooper plates, is provided by a pulsed valve with an opening time of $\sim$3 ms.
An adjustable aperture was placed before the lens to adjust the laser energy and the beam size.

The spectrum of HHs was measured by a lab-built soft X-ray spectrometer that contains a $\sim$ 100-$\mu$m-wide slit, a 1200 grooves/mm (for Ar gas) or 2400 grooves/mm (for Ne and He gas) aberration-corrected grating (Hitachi), a micro-channel plate (MCP) with a phosphor screen detector that is 2.5 m away from the gas cell, and a CCD camera for capturing the image on the phosphor screen.
The photon energy axis of spectrum was calibrated using absorption edges of Al(72.8 eV), Cl (200 eV) and plasma emission lines (307.9 eV, 354.6 eV) of Carbon.
To correctly measure the pulse energy of the HHs, a 200-nm-thick Ti film, which was tested to be strong enough to block the MIR laser pulse without damage, was employed before a calibrated photodiode (AXUV100, IRD). 
From spectral data and the pulse energy of the HHs after the Ti film, we can calibrate our spectrometer to measure both the spectrum and pulse energy simultaneously.

%


\subsection{Optimization of HHG driven by a loosely focused TW MIR pulse}

\begin{figure}[htbp]
\centering
\includegraphics[width=\linewidth]{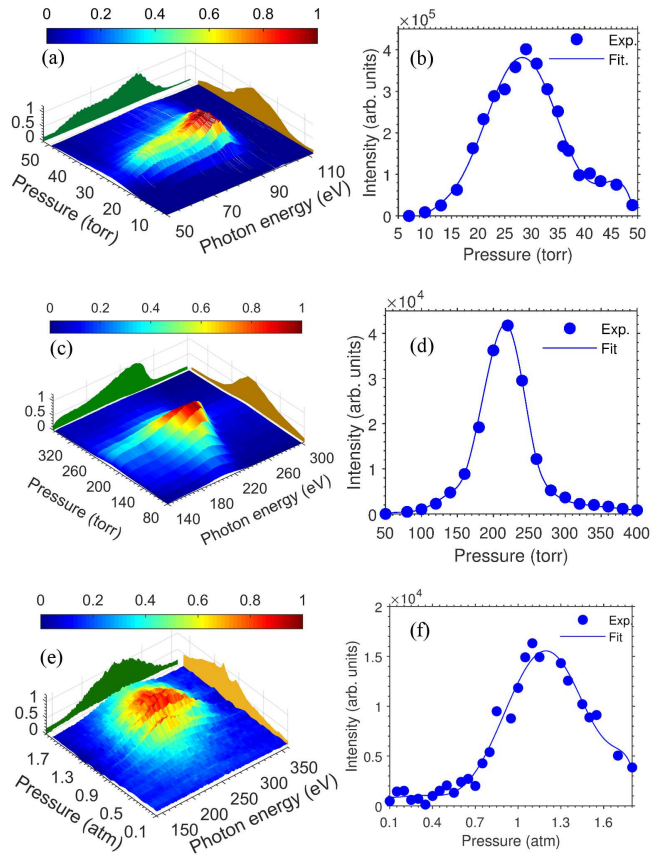}
\caption{
Pressure-dependent HHs spectra for Ar (a), Ne (c), and He (e), respectively. (b), (d) and (f) show the pressure-dependent yield for 90 eV from Ar, 240 eV from Ne, and 300 eV from He, respectively.
 The solid curves in (b), (d) and (f) are polynomial fits to the data.
The laser intensities were 1.2$\times$10\textsuperscript{14} W/cm\textsuperscript{2}, 3.9$\times$10\textsuperscript{14} W/cm\textsuperscript{2}, and 5$\times$10\textsuperscript{14} W/cm\textsuperscript{2} for Ar, Ne, and He respectively, as estimated by the cutoff energy.}
\label{PressDepend}
\end{figure}

We began the experiment by optimizing the pressures of Ar and Ne for generating HHs in the 4-cm-long gas cell.
The target gas was statically filled in the long gas cell.
The results are presented in Figs. \ref{PressDepend} (a) and (c), respectively.
The optimized pressures were approximately 30 torr (0.039 atm) and 220 torr (0.29 atm) for Ar and Ne, respectively, which are around one order lower than those in other reported studies \cite{TeichmannNC,ChenPRL}. Figures \ref{PressDepend} (b) and (d) present the pressure-dependent HH yield at each photon energy for Ar and Ne, respectively. In Ar, the optimized photon energy was near 90 eV with the highest yield. In Ne, the best optimized photon energy was near 240 eV, and the yield was more sensitive to pressure. To further increase the photon energy, we changed the gas to He. 
However, the highest statically filled pressure of He is approximately 300 torr, which is lower than the optimized pressure, due to fast leakage and the limited pumping capacity. 
Thus, we  introduced the high pressure gas cell which consists of the double cell structure.
The gas to the new designed inner cell, which was 3-cm-long, was provided by a pulsed gas valve. 
Thus, we could significantly reduce the gas load to the pumping systems while keeping sufficient pressure in the interaction region. 
The vacuum level in the target chamber was maintained to be better than 0.01 torr. The result is presented in Fig. \ref{PressDepend} (e), which indicates an optimized pressure of $\sim$1.2 atm.
Note that the pressure was calibrated to a continuous backing pressure by monitoring the spectral blue shift \cite{SpecBlueShift} of the driving laser after the gas cell. 
We measured spectral blue shift of driving laser after gas medium when the gas cell was filled with static gas and pulsed gas. By comparing spectrum blue shift amount in both cases, we calibrate the pulsed pressure to static baking pressure, which is very helpful for comparison and discussions.
Figure \ref{PressDepend} (f) presents the HH yield of 300 eV that is dependent on the He pressure. 

\begin{figure}[h]
\centering
\includegraphics[width=\linewidth]{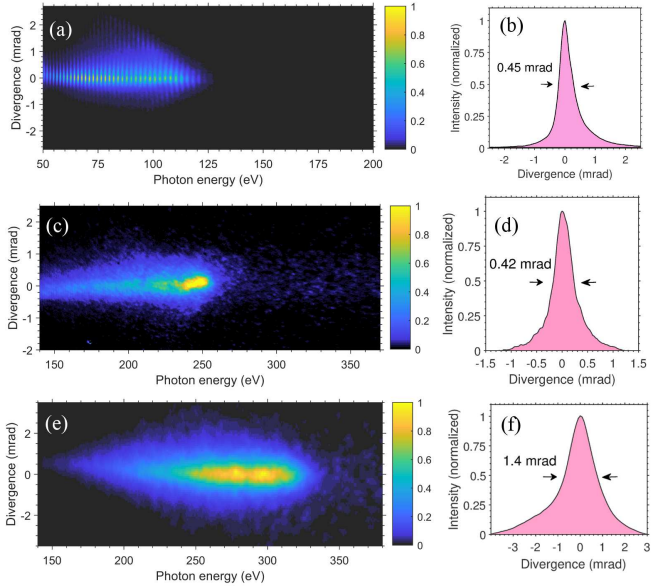}
\caption{
Experimental spectra showing both the photon energy (horizontal axis) and divergence (vertical axis) in different gases: (a) Ar, (c) Ne, and (e) He. (b), (d), and (f) present the beam divergence profiles for 60-80 eV in (a), 240-260 eV in (c), and 290-310 eV in (e), respectively.}
\label{Divergence}
\end{figure}

Figure \ref{Divergence} shows the two-dimensional spectra in different gas media.
The spectrum obtained in Ar is presented in Fig. \ref{Divergence} (a), which indicates a very small beam divergence across the entire spectral range.
The beam divergence values are calculated based on beam size of HHs measured on the detector of spectrometer and propagation distance from the generation point, since there is no optics that change beam divergence during its propagation.
The divergence profile for 60-80 eV is plotted in Fig. \ref{Divergence}(b) with a divergence value of 0.45 mrad (FWHM).
Figure \ref{Divergence}(c) presents the spectrum for Ne.
The profile of divergence for 200-220 eV is presented in Fig. \ref{Divergence}(d) with a divergence value of 0.42 mrad (FWHM).
Then, we checked the spectral divergence in He gas, which is presented in Fig. \ref{Divergence}(e).
Figure \ref{Divergence}(f) presents the divergence profile for 260-280 eV with a value of 1.4 mrad (FWHM).
The divergence profiles in Fig.\ref{Divergence}(b,d,f) are Gaussian-like with small divergences, indicating good phase matching along the pump propagating direction \cite{TakAPL}. Good beam profiles for the soft X-ray pulses are essential for future applications.

\begin{figure}[t]
\centering
\includegraphics[width=\linewidth]{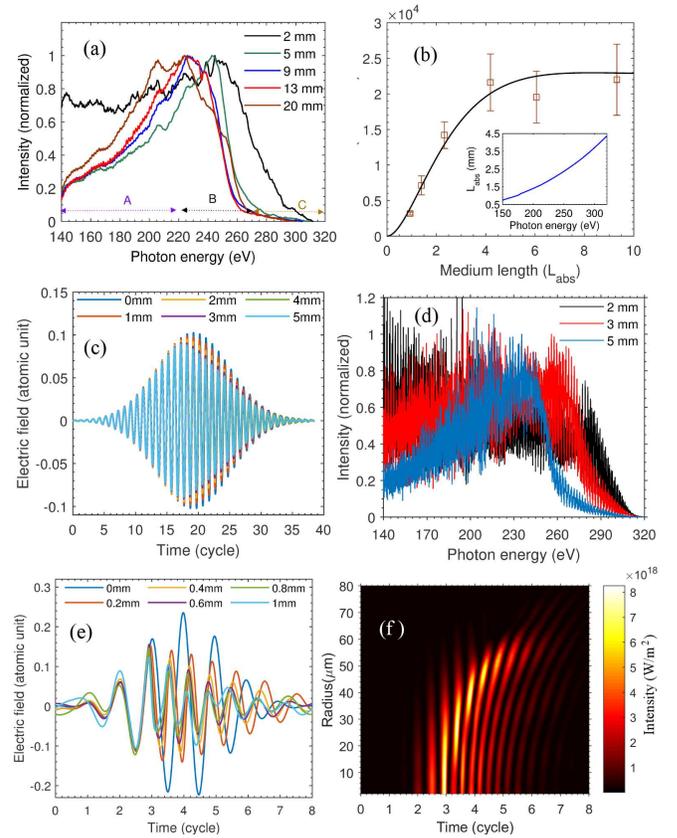}
\caption{
(a) Normalized soft X-ray spectra observed under different Ne gas lengths. The gas pressure was maintained at 205 torr. (b) The yield (square dots with error bars) at 240 eV under different Ne gas lengths. The black solid curve indicates the calculated result when the coherence length (L\textsubscript{coh}) was 13 times the absorption length (L\textsubscript{abs}) using formula (1) in Ref. \cite{ConstantPRL}. The laser intensity was 3.9 $\times$ 10\textsuperscript{14} W/cm\textsuperscript{2}. The inset indicates the calculated absorption length for each photon energy in 205 torr of Ne gas.
(c) Electric field after different propagation lengths. (d) Simulated HHs spectra after different medium lengths. (e) Electric field waveform after propagation for a few-cycle waveform using the ``over drive'' method. (f) Spatial and temporal profiles of the driving laser field after 0.5-mm-long medium using the ``over driven'' method.}
\label{Simulation}
\end{figure}

In the next experiment, we determined our procedures to optimize the HH yield under different medium lengths to ensure that our experiment is optimized for absorption-limited conditions.
With Ne gas as an example, the spectra are presented in Fig. \ref{Simulation} (a).
The intensity of each harmonic was normalized by the maximum intensity.
 When the medium was 2 mm, the spectrum exhibited a plateau structure at 140-250 eV, followed by a cutoff.
This spectral feature is similar to a single-atom response, which indicates that propagation did not yet play a significant role. When the medium length increases, propagation becomes important, which includes effects of both phase matching and absorption of soft X-rays.
For example, the spectral shape for Ne lengths over 5 mm can be classified into three regions, which are indicated by A, B, and C in Fig. \ref{Simulation} (a). 
Region A shows an absorption-dominated feature because the absorption decreases with photon energy as shown by the absorption lengths in the inset of Fig. \ref{Simulation} (b).
In region B, the slope of the spectrum becomes steeper compared with that of the 2-mm length.
Ionization-limited phase matching is believed to be the main reason for this.
Under our experimental conditions, the ionization ratio needs to be less than 0.25\% to fulfill phase matching, which limits the maximum phase-matched photon energy of $\sim$ 260 eV.
This is consistent with the photon energy in region B, which was thus the ionization-limited phase-matched cutoff energy.
In region C, ionization was greater than 0.25\%, and phase matching could not be satisfied.
Note that this phenomena occurs in temporal and spatial domains, both of which have intensity distributions.
Thus, by shortening the pulse duration or by mixing other laser wavelengths to reshape the waveform, the ionization-limited phase-matched cutoff energy can be extended due to reduced ionization.
In Fig. \ref{Simulation} (b), we plotted the yield at 240 eV as a function of the medium length.
By fitting the data using formula (1) in Ref. \cite{ConstantPRL}, we could confirm that the coherence length (L\textsubscript{coh}) was over ten times the absorption length (L\textsubscript{abs}), which satisfies well the absorption-limited condition.
The yield saturated when L\textsubscript{coh} $\geqslant$ 9.3 L\textsubscript{abs} (20 mm). Thus, the length and pressure of the Ne gas was set to 20 mm and near 205 torr, respectively.

Another feature observed in Fig. \ref{Simulation} (a) is that the highest yield of the HH shifted to a lower photon energy when the medium length increased.
To determine the reason, we performed a simulation that considered the propagation effect.
In our simulation, the HHG process is modeled by the single-atom response, the propagation of the fundamental driving laser, and the harmonic field in the gas cell. 
We applied the strong-field approximation (SFA) \cite{Lewenstein} model to calculate the harmonic radiation at each plane by inputting the spatiotemporal profile of the IR pulse from the propagation simulations. 
The propagation of the IR driving laser was simulated by solving a three-dimensional (3D) Maxwell wave equation in cylindrical coordinates, which included the effects of refraction, absorption, and nonlinear polarization (see Ref. \cite{Lan2010}). 
The propagation of harmonic radiation is also described by Maxwell's wave equation, which contains both linear and nonlinear polarization terms. 
After the propagation in the medium, the radiation from each plane was integrated to obtain the macroscopic harmonic field at the exit of the cell. 
All parameters used in the simulation were found to be consistent with that of the experiments. 
Figure \ref{Simulation}(d) presents the same trend as in Fig. \ref{Simulation} (a), which can be explained by two reasons as determined for our simulation.
One reason is the laser intensity distribution in the Rayleigh range along the propagation, which is higher near the focus but lower when moved away from the focus.
The HHs with the highest cutoff are generated only at the highest laser intensity, followed by absorption during propagation.
The other reason is the electric field modulation of the driving laser during propagation, which is presented in Fig. \ref{Simulation}(c).
Even though we maintained a very low ionization rate, the electric field strength slightly decreased during the long propagation length.
Thus, the cutoff energy decreased when a longer medium length was employed.

\subsection{Comparison of a loosely focusing geometry and an over driven scheme}

For comparison, we also performed a simulation of the $over~driven$ scheme for a few cycles of the laser in a Ne gas medium at a 1.5-atm pressure. 
The laser at 1.8 $\mu$m with the two-cycle pulse duration (FWHM) was focused on 1 mm after a 1-mm-thick gas medium with a very high intensity of 2 $\times$ 10\textsuperscript{15} W/cm\textsuperscript{2}. Both temporal and spatial profiles of the laser were perfect Gaussians.
From the calculation, the electric field in the temporal domain was significantly modulated with a very short propagation distance, as presented in Fig. \ref{Simulation}(e).
Spatial modulation is also significant, as presented in Fig. \ref{Simulation}(f), which was not obvious in our experiment according to the simulation.
These characteristics require an excellent driving laser spatial profile for the $over~driven$ scheme to avoid beam breakup \cite{JohnsonSciAdv}.
The experiment under the $over~driven$ needs to be optimized to a critical condition in which the laser intensity distribution along the propagation direction is nearly constant in a region (clamped by plasma defocusing) that matches the medium length. 
This clamped region is much smaller than the Rayleigh length and cannot be simply calculated.
Based on these considerations, the $over~driven$ scheme may have difficulty in terms of energy scaling of HHs using a high-energy driving laser, e.g., 100 mJ with TW peak power, which generally has a worse beam quality with hot spots and a higher pulse energy fluctuation.
Moreover, the atto-chirp cannot be easily predicted, which is inconvenient for the pre-design of chirp compensation, due to the strong distortion of the driving laser field during propagation in the medium.
In our method, energy scaling of HHs is straightforward without worrying about beam breakup because a much lower gas pressure is employed under significantly weaker ionization (0.2\% level).
The atto-chirp and Rayleigh range can be easily predicted.

\subsection{Energy scaling of water window HHG}

\begin{figure}[t]
\centering
\includegraphics[width=\linewidth]{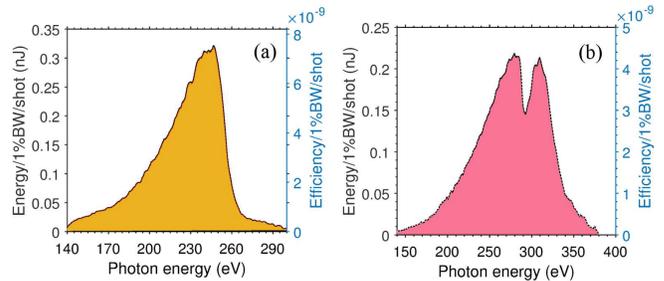}
\caption{
Energy and conversion efficiency per shot in 1\% bandwidth of soft X-rays generated from the Ne (a) and He (b) gas cells.}
\label{HHenergy}
\end{figure}

Next, we characterized the pulse energies of the soft X-ray HHs.
A 200-nm-thick titanium filter, whose transmission was measured to be ~15\% at 210 eV, was employed to block the MIR driving laser.
 Then, the pulse energy of the soft X-rays was measured using a calibrated photo diode (AXUV100, IRD). 
 The spectrum was recorded by the spectrometer.
 Based on both the pulse energy and spectrum, we calibrated our spectrometer for both energy and spectral measurements.
 Figure \ref{HHenergy}(a) presents the pulse energy and conversion efficiency for 1\% bandwidth (1\%BW) of soft X-rays obtained in Ne gas.
 The pulse energy near 250 eV in 1\% BW was 0.32 nJ with a conversion efficiency of 7.4 $\times$ 10\textsuperscript{-9}.
 In the photon energy range of 140-300 eV, the total pulse energy was approximately 10 nJ, indicating a conversion efficiency of 2.3 $\times$ 10\textsuperscript{-7}. 
 The maximum photon energy was around 300 eV, which just enters the water window region.
To further increase the photon energy deep into the water window region, He gas was employed in a pulsed gas cell as described above.
Figure \ref{HHenergy}(b) presents the soft X-ray spectrum with calibrated energy and conversion in 1\% BW per shot in He. 
At 310 eV, the pulse energy was 0.22 $\pm$ 0.01 nJ in 1\% BW, indicating a conversion efficiency of 4.4 $\times$ 10\textsuperscript{-9}.
For the water window soft X-rays at 284-380 eV, the total pulse energy was 3.53 $\pm$ 0.12 nJ. The total conversion efficiency of the water window soft X-rays was (7.20 $\pm$ 0.24) $\times$ 10\textsuperscript{-8}.
The photon flux in the water window region was (7.12 $\pm$ 0.30)$\times$10\textsuperscript{8}/s.
Note He has a low absorption for soft X-rays, e.g., in the water window and beyond. For higher photon energies, the absorption is even lower. 
Thus, He is a good gas medium for generating high-energy HH photons. 
In our experiment, the absorption length of He at 310 eV under an interaction pressure of 1.2 atm was $\sim$ 20 mm. 
Note that 1.2 atm in our experiment was the backing pressure, which should be higher than the real pressure in the interaction region. 
Thus, the absorption-limited condition requires that the He gas medium length be over 60 mm ($L_{\rm med} > 3 L_{\rm abs}$). 
However, the Rayleigh length of our laser limited our gas medium length to $\sim$ 30 mm, which was less than half of the optimized length. 
In principle, by employing a guiding scheme to extend the interaction length to 100 mm in our experiment, the pulse energy of the HHs in the water window region could be increased to over 10 nJ.

\subsection{NEXAFS by the nano-joule water window beam}

\begin{figure}[h]
\centering
\includegraphics[width=\linewidth]{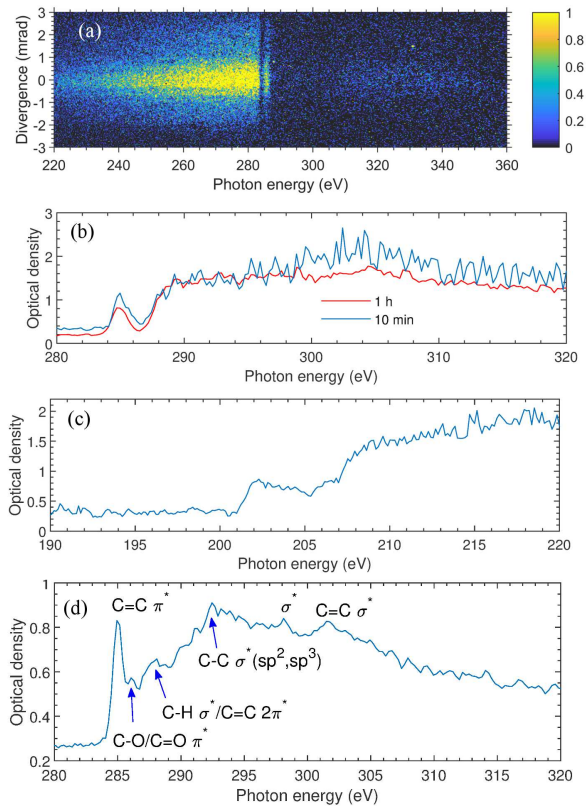}
\caption{
(a) 2D Spectrum after passing through a 1-$\mu$m-thick Mylar film. (b) Absorption spectrum near the carbon K-edge in (a). (c) Absorption spectrum near the L-edge of chlorine after a 1.2-$\mu$m-thick Parylene-D film.
(d) Absorption spectrum near the carbon K-edge after a 0.25-$\mu$m-thick Parylene-C film.
}
\label{XAFS}
\end{figure}

One application of our soft X-ray harmonics is near edge X-ray absorption fine structure spectroscopy.
To improve the spectral resolution of our spectrometer, we employed an X-ray CCD as the detector for the spectrometer with a pixel size of 13 $\mu$m with an array of 2048$\times$2048. 
The spectral resolution was estimated to be better than 0.5 eV near the carbon K-edge.
The first sample we used was a 1-$\mu$m-thick Mylar film.
The transmitted spectrum is presented in Fig. \ref{XAFS} (a) in which fine absorption structures near the carbon K-edge can be clearly observed. 
This image is obtained by the accumulation of 6000 shots (10 min).
Blue line profile in Fig. \ref{XAFS} (b) presents the absorption spectrum evaluated using spectra with and without the Mylar film.
The clear absorption peak near 285 eV resulted from the 1s to $\rightarrow$ $\pi$\textsuperscript{*} transition of the C=C bond of the benzene ring \cite{PopmintchevPRL}.
Because the signal, where the  photon energy is higher than the carbon K-edge, is strongly absorbed by a 1-$\mu$m-thick Mylar film, the signal-to-noise ratio becomes a poor. 
By increasing a shot number, we can improve the the signal-to-noise ratio and get a smooth data, which is shown by the red profile.
There was no obvious fine structures observed above 290 eV even though the measurement time was extended to 1 hour (36000 shots).
For a 1-$\mu$m-thick Mylar film, the accumulation of 6000 shots was enough to distinguish the fine structure.
Then, we changed the sample to a 1.2-$\mu$m-thick Parylene-D film.
The fine structure near the carbon K-edge cannot be observed in the transmission spectrum even we accumulated 36000 shots, which is because the sample was too thick for the penetration depth of soft X-rays.
Instead, a fine structure can be observed near 202 eV, where the absorption spectrum is plotted in Fig. \ref{XAFS} (c). The absorption peaks at 202, 206, and 208 eV result from L-edge transitions of chlorine \cite{KawerkCL}.
In fact, the penetration depth near 300 eV was only $\sim$ 0.2 $\mu$m for the Mylar and Parylene films. The above two films were too thick for this experiment.
To avoid absorption saturation and observe more absorption features, we employed a thinner sample, which was a 0.25-$\mu$m-thick Parylene-C film.
The NEXAFS result is presented in Fig. \ref{XAFS} (d).
Many fine structures can be observed near the carbon K-edge in the absorption spectrum of Fig. \ref{XAFS} (d) with absorption peaks near 285, 288, 292, 298, and 302 eV, which were assigned to transitions from 1s to C=C $\pi$\textsuperscript{*}, C-H $\sigma$\textsuperscript{*}/C=C 2$\pi$\textsuperscript{*}, C-C $\sigma$\textsuperscript{*} (sp\textsuperscript{2},sp\textsuperscript{3}), $\sigma$\textsuperscript{*}, and C=C $\sigma$\textsuperscript{*}, respectively \cite{Kikuma,WachulakOE}. Concerning the absorption peak at 286 eV, it was assigned to a transition from 1s to $\pi$\textsuperscript{*} of C-O or C=O bonds, which resulted from photo-oxidation or oxygen contamination of the Parylene-C film \cite{Pruden}.

\section{Conclusion and outlook}

Taking our measured NEXAFS results into account, we discuss the possibility of a single-shot XAFS experiment.
Realizing a single-shot measurement technique is very important for not only tracking irreversible phenomena but also saving a measurement time in experiments.
In our NEXAFS experiment, we could observe fine absorption structures near Carbon k-edge in 10 min for 1-$\mu$m-thick Mylar and 2 min for 0.25-$\mu$m-thick Parylene-C films, when the beam diameter of water window HH is around 1 cm  and a slit width of our spectrometer is 100 $\mu$m. 
With our current measurement system, only 2\% harmonic energy is utilized to take an absorption spectrum and 98 \% of the energy is lost at the spectrometer slit.
By employing a toroidal mirror to collect all water window HHs energy for spectroscopy or imaging, the measurement time will be shortened by $\sim$ 100 times.
Therefore, our nano-joule water window harmonic source enables us to take an absorption spectrum within a few seconds.
Of course, if the sample thickness is thinner, the current output flux will enable single-shot measurements.
Moreover, our robust energy scaling method on HHG, which generated a nano-joule level water window soft X-ray, has the advantage that the harmonic output yield can be further scaled up by increasing the harmonic emission size.
Thus, by increasing pumping energy to 100 mJ with a deformable mirror to improve its wave front \cite{TakAPL}, we will be able to perform single-shot NEXAFS experiments on a thicker sample.

In addition, we evaluate the possibility of single-shot live-cell imaging by contact microscopy using our nano-joule water window HHs. 
Thus far, the most practical source of x rays has been synchrotron radiation produced by an accelerator. 
Images of live cells, however, cannot be captured without freezing the sample even using the world's largest synchrotron radiation facility due to its low instantaneous power.
Currently, XFEL facility can provide a probe pulse for imaging live-cell with X-ray diffraction technique \cite{kim_xray}.
Single-shot images of live cells are expected to provide valuable information that is lost when freezing samples.
According to a recent work using a plasma light source \cite{Ayele2017}, photon flux of 1.9$\times$10\textsuperscript{5} photons/$\mu$m\textsuperscript{2}/pulse will be able to perform cell imaging in a single shot manner. 
Thus, by focusing our nano-joule water window HHs to a 10 $\mu$m spot size at 310 eV with a 30 eV bandwidth, single-shot cell imaging by contact microscopy can be performed, which is very helpful for live cell imaging with sub-100 nm resolution.
The developed table coherent water window HH source is feasible to demonstrate single-shot live-cell imaging.

\begin{figure}[h]
\centering
\includegraphics[width=\linewidth]{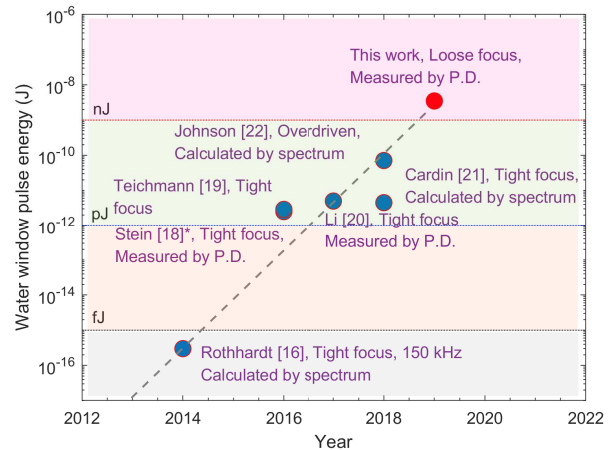}
\caption{
Pulse energies of water window soft X-ray HHs that have been reported in recent years. P.D., photodiode. * Estimated according to the P.D. measured energy and spectrum reported in each paper.}
\label{WaterWindowEnergy}
\end{figure}

In this study, even though we employed more than one order lower pressure, we achieved a high conversion efficiency using simulation predictions \cite{KartnerEffScaling}.
 Our demonstration is very helpful for further energy scaling of soft X-ray HHs using even longer wavelengths.
Owing to the much lower target gas pressure, we can avoid safety issues and technical challenges when maintaining sub-100-atm gas pressures.
Compared with the $over~driven$ method for HHG, which requires an excellent laser quality and has difficulty achieving energy scaling, our method is straightforward for energy scaling with a reduced beam quality, which is especially important for high-energy lasers, e.g., 100 mJ with a TW class peak power.
Additionally, less temporal distortion to both the driving laser and soft X-rays is expected when using our method under low gas pressures and a weakly ionized medium.
 This method will be very helpful for generating temporal profile-predictable IAPs; thus, chirp compensation for IAPs can be pre-designed. Moreover, we can avoid possible incoherent emissions, resulting from electrons that are recolliding with adjacent atoms in high-pressure cases \cite{MarcusHighPressure}, especially when a significantly longer electron trajectory occurs because of the use of a longer wavelength laser.
To clearly illustrate our achievement, we plot the energy of water window harmonics that have been reported in recent years in Fig. \ref{WaterWindowEnergy}.
Our methods achieve the highest pulse energy for water window region HHs owing to the much higher driving laser energy.
We found that smaller laser beam size selected by aperture gives a better focus quality as well as HHG signal intensity and profile. This means that outer side of the laser beam does not have a good wave front, which is similar to our previous work \cite{TakAPL}. 
Thus, we will improve the wave front of the driving laser using a deformable mirror system in which the conversion efficiency of our soft X-ray HHs will be further improved by several times \cite{TakAPL}. 
Additionally, by increasing the medium length to reach the absorption limited condition, we can further increase the energy of HHs by another two to three times.

In summary, by employing a TW class MIR fs laser to drive HHG, we have obtained nJ-class soft X-rays up to the water window region.
The pulse energy reaches 10 nJ for 150-300 eV and 3.53 nJ in the water window, which will be approximately two orders higher than those in reported studies, even though we employ an approximately one order lower gas pressure.
The low gas density significantly prevents spatial and temporal distortion of the driving laser and HHs.
Moreover, we observed the ionization-limited cutoff energy, which is a direct evidence that phase matching is more sensitive to plasma dispersion (requires very low ionization rate) in MIR fs laser-driven HHG.
Our work establishes a universal method for energy scaling of soft X-ray HHs.
We will further increase the pulse energy in the water window region to 10 nJ by improving the driving laser wavefront and extending the medium length.
Furthermore, our scheme is straight forward for energy scaling. For example, increasing pulse energy to 200 mJ and focal length to 4 m, the water window HHs energy will be increased by 4 times.
This universal energy scalability is extemely important to develop the high-energy water window HH source.
As the first application of our harmonic sources, fine absorption spectral structures were observed for different samples near the K- and L-edges.
Our developed HH source enables us to perform time-resolved nonlinear NEXAFS using strong HHs in an attosecond-pulse-pump and attosecond-pulse-probe manner \cite{Chen2001}.
The systematic study of a robust energy scaling method on HHG opens the door for demonstrating of single-shot XAFS and live-cell imaging with a femtosecond time resolution.
Moreover, our high-energy soft X-ray pulses will be very useful in applications of XFEL seeding, nanolithography, ultrafast dynamics studies, and nonlinear soft X-ray physics.

\begin{acknowledgments}
The authors would like to acknowledge Dr. Tomoya Okino from RIKEN, Dr. Nobuaki Kikuchi from Tohoku University, and Dr. Katsuya Oguri from NTT Basic Research Laboratories.

This work was supported in part by the Ministry of Education, Culture, Sports, Science and Technology of Japan (MEXT) through Grant-in-Aid under Grant 17H01067, in part by the MEXT Quantum Leap Flagship Program (Q-LEAP) Grant Number JP-MXS0118068681, in part by the FY 2019 President discretionary funds of RIKEN, and in part by the Matsuo Foundation 2018. 
K. M. acknowledges support by MEXT through Grant-in-Aid under Grant 19H05628.
P.F.L. acknowledges support by National Key Research and Development Program (2017YFE0116600) and National Natural Science Foundation of China (91950202).
\end{acknowledgments}

\vspace{2mm}

\noindent
Y.F. current address: Xi’an Institute of Optics and Precision Mechanics, Chinese Academy of Sciences, Xi'an, Shaanxi 710119, P.R.China.



\end{document}